# Algorithm-based diagnostic application for diabetic retinopathy detection

Agnieszka Cisek*[1], Karolina Korycińska[1], Leszek Pyziak[2], Marzena Malicka[2], Tomasz Więcek[2], Grzegorz Gruzeł[3], Kamil Szmuc[4], Józef Cebulski[3], Mariusz Spyra[1]

* Corresponding author: aga.cisek@wp.pl

[1] Visum Clinic, Litewska St. 4C, 35-302 Rzeszow, Poland

[2] Depertment of Applied Optics, The Faculty of Mathematics and Applied Physics, Rzeszow University of Technology, Av. Powstańców Warszawy 8, 35-959 Rzeszow, Poland

[3] Department of Optometry and Spectroscopy, Institute of Physics, College of Natural Sciences, Profesora Stanisława Pigonia St. 1, 35-310 Rzeszow, Poland

[4] Institute of Materials Engineering, College of Natural Sciences, University of Rzeszow, Profesora Stanisława Pigonia St. 1, 35-310 Rzeszow, Poland

**Abstract**

Diabetic retinopathy (DR) is a growing health problem worldwide and is a leading cause of visual impairment and blindness, especially among working people aged 20-65. Its incidence is increasing along with the number of diabetes cases, and it is more common in developed countries than in developing countries. Recent research in the field of diabetic retinopathy diagnosis is using advanced technologies, such as analysis of images obtained by ophthalmoscopy. Automatic methods for analyzing eye images based on neural networks, deep learning and image analysis algorithms can improve the efficiency of diagnosis. This paper describes an automatic DR diagnosis method that includes processing and analysis of ophthalmoscopic images of the eye. It uses morphological algorithms to identify the optic disc and lesions characteristic of DR, such as microaneurysms, hemorrhages and exudates. Automated DR diagnosis has the potential to improve the efficiency of early detection of this disease and contribute to reducing the number of cases of diabetes-related visual impairment. The final step was to create an application with a graphical user interface that allowed retinal images taken at cooperating ophthalmology offices to be uploaded to the server. These images were then analyzed using a developed algorithm to make a diagnosis.

**Introduction**

Diabetic retinopathy (DR), which is one of the secondary complications of diabetes, is a global health problem with an increasing incidence. It poses a serious threat to vision and is now a major healthcare challenge. It is the leading cause of visual impairment and blindness worldwide, and its incidence is increasing as the number of diabetes cases increases [1]-[3] Diabetic retinopathy is the leading cause of blindness and visual impairment in the working population (people aged 20-65) worldwide [4]-[6]. The total number of people with blindness due to hyperglycemia accounts for 2.6% of the global population [1]. In developing countries, diabetic retinopathy is a less common condition (19.9%) compared to developed European countries (45.7%). Within South Asia, depending on diet patterns and lifestyle differences, urban populations are more prone to developing DR than suburban or rural communities [7]. Some studies prove that DR is more common in young people with type 1 diabetes than type 2, and therefore represents a significant burden on the socioeconomy, as it largely affects working-age people [1]. The prevalence of DR varies by population, with some ethnic groups showing a higher risk[8]. Several risk factors have been identified, including duration of diabetes, lack of glycemic control, hypertension, dyslipidemia and genetic susceptibility [9]-[11]. Persistently high blood glucose levels contribute to oxidative stress and inflammation, leading to cellular damage within the retina [12]. Disturbances in insulin signaling processes also contribute to the pathogenesis of diabetic retinopathy [13]. The exact mechanisms underlying the development of diabetic retinopathy are complex and involve multiple molecular pathways, including changes in angiogenesis, the onset of inflammation and neurodegeneration [14]. Early detection and appropriate prevention are key to preventing or delaying the progression of DR and reducing the incidence of visual impairment in diabetes [15], [16]. Most often, microaneurysms (MA) are visible in the retina as the first sign of DR. However, there may be more retinal lesions. Indications of DR further include exudates (EX) and hemorrhages (HM), as well as abnormal growth of blood vessels. DR usually has two stages called non-proliferative DR (NPDR) and proliferative DR (PDR) retinopathy [17].

In the first stage of the disease (NPDR), fluid leaking from damaged blood vessels accumulates on the retina. The retina is then moist and swollen. At this stage, there are various signs of retinopathy, for example, HM, MA, EX, and interretinal vascular abnormalities (IRMA). In the further course of the disease (PDR), new blood vessels appear, which are even more susceptible to damage [18].

Untreated diabetic retinopathy progresses, so early detection is very important. Early discovery and treatment of DR play a major role in preventing adverse effects such as blindness. People at high risk of occurrence should be screened by taking and analyzing pictures of the retina. Stationary funduscopes or portable ophthalmoscopes are used to take such images. The disadvantages of the former are the cost of their purchase and their stationary nature - it is the patient who has to come to a specialized clinic to have the appropriate picture taken. However, the images taken are usually of very good quality, which is not so obvious with portable devices. Taking photos "hand-held" can result in blurring of the photo, recording the wrong area of the retina, its displacement or uneven illumination, caused by external light entering the camera. Handheld devices, however, are cheaper to purchase, so smaller clinics can afford them. Another advantage is that they are portable, making them easier for the patient to use [19].

In recent years, advanced technologies have been developed to analyze images acquired by ophthalmoscopy. These methods are based on the analysis of the shape, size and texture of retinal structures, which allows more precise identification of pathological changes. The use of ophthalmoscope image analysis can significantly improve the efficiency of diabetic retinopathy diagnosis and allow early detection of lesions that are difficult to see with traditional ophthalmoscopic examination. Recently proposed methods for automatic or semi-automatic computer-based analysis of retinal images are mainly based on neural networks [20], [21], deep learning [22], [23] and image analysis algorithms [24], [25] . With the expansion of available image databases, there is growing interest in the implementation of automatic support tools in the diagnostic process [26]-[28].

The purpose of this study is to develop an automated method for the diagnosis of diabetic retinopathy using the following: an algorithm for pre-processing (normalization) and analysis and qualitative selection of images derived from a hand-held ophthalmoscope; an algorithm for finding characteristic elements of the retina; an algorithm for detecting lesions characteristic of diabetic retinopathy (microaneurysms, hemorrhages, hard exudates and soft exudates).

**Methodology**

In this study, we analyzed the results of retinal images obtained using the Optomed Smartscope® PRO ophthalmoscope. This is a device which allows image registration, without the need to sprinkle the patient's eye. The resolution of the camera's matrix is 1.77 Mpix, with a maximum image resolution of 1536 by 1152 pixels. The device has a wide viewing angle of 40 degrees, which makes it possible to record in a single image, the entire area necessary for a correct diagnosis. The work analyzed 647 retinal images of patients participating in the study.

All provided images are cropped to 90% of the length of their original radius. Morphological algorithms were used to identify the optic nerve disc, including the dilatation algorithm and the erosion algorithm defined in the OpenCV library.

**Results and discussion**

The photos obtained during the research and used as input material, necessary for building the algorithm, have very different quality. Unfortunately, the following defects are visible in a large number of them: incorrect photo geometry, out-of-focus photo, uneven illumination of the photo (overexposed or underexposed photo), artifacts visible in the field of view (eyelids, eyelashes, rainbow effects). Examples of photos showing sample imaging errors are presented below (Figure 1).

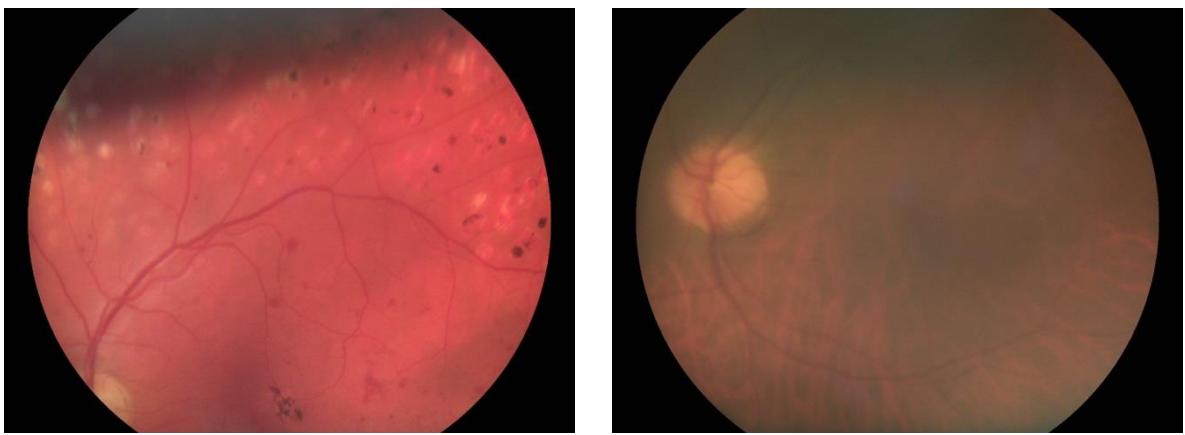

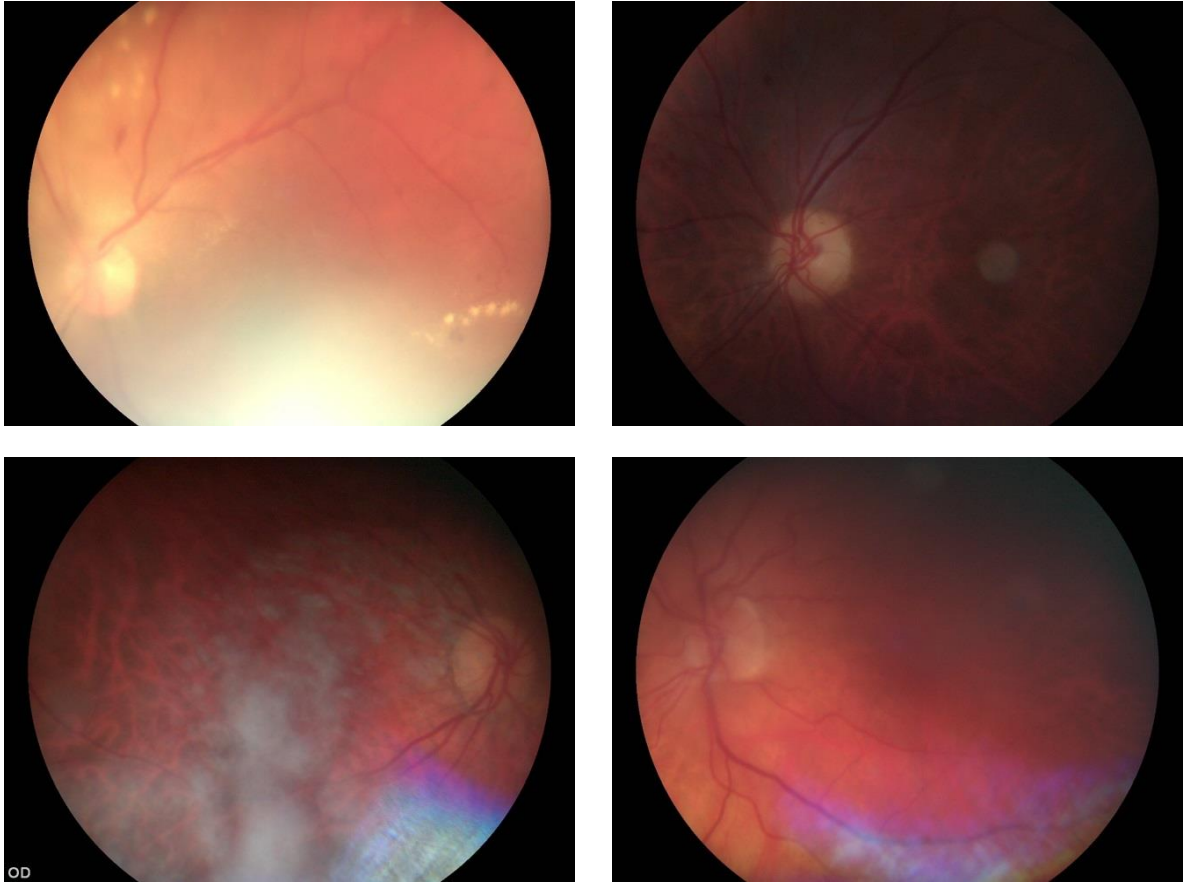

*Figure 1. Retinal images with examples of defects making them impossible to be correctly interpreted by the algorithm being developed.*

Defects visible in the recorded images are mainly the result of errors made during the taking of the picture. Their presence makes it much more difficult or even impossible to correctly analyze and find signs of diabetic retinopathy. As part of the implementation of the research work, it was decided that the quality should be improved, the area affected by defects should be cropped if possible, or the photo should be eliminated from further analysis. Figure 2 shows examples of retinal images, which were classified as good quality images, possible to analyze using the developed algorithm.

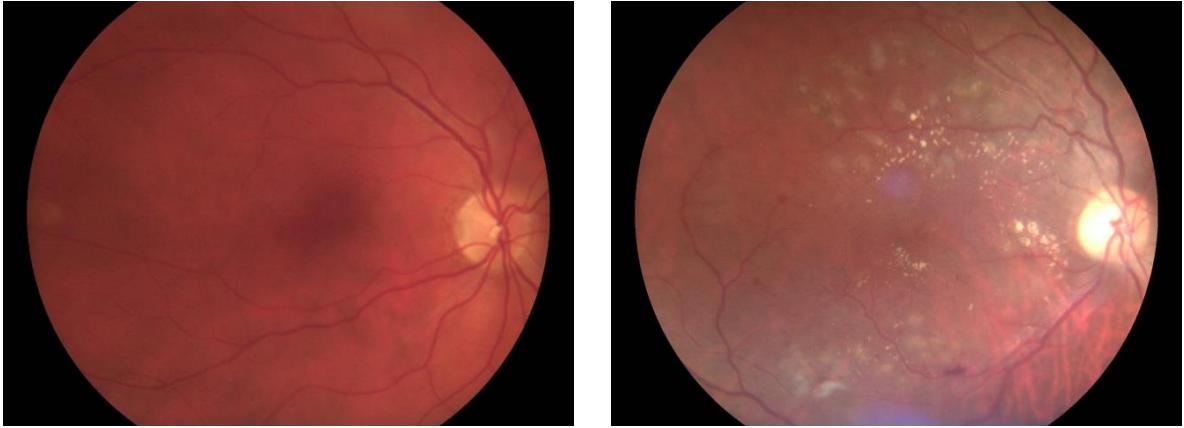

*Figure 2: Examples of retinal images classified as good quality images.*

The figure below shows the distribution of light intensity in the original image sample of the patient's retina, as well as the three different components of this image i.e. the red, green and blue channels (Figure 3).

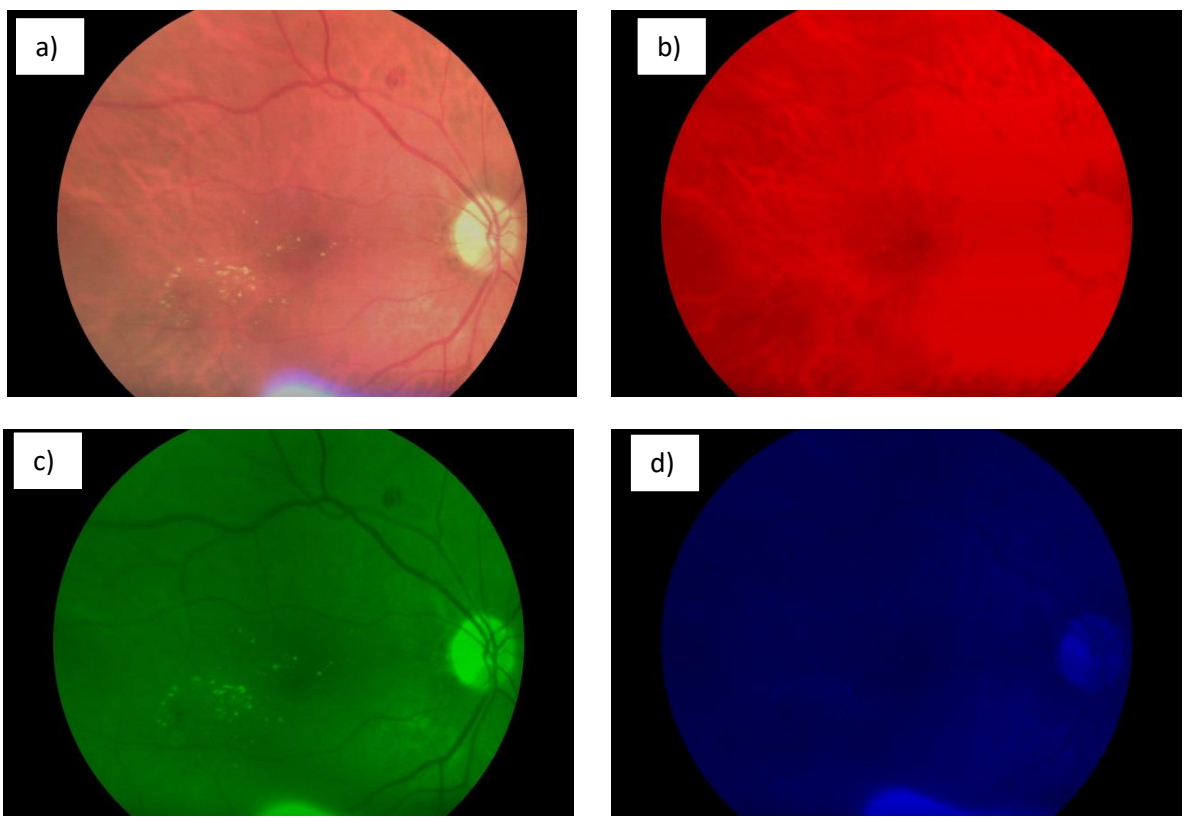

*Figure 3. Photos showing the distribution of light intensity in: a) the original intensity distribution; b) the red channel; c) the green channel; d) the blue channel.*

The brightening visible in the lower area of the image is due to incorrectly taken images and was not considered in the analysis using the algorithm.

The image (Figure 3) shows a large difference in the information content of the different channels. The green channel (Figure 3c) contains the most information. The retinal objects visible in it are characterized by high contrast. Both anatomical objects of the eye and lesions are well visible. The red channel (Figure 3b), due to the large number of blood vessels in the retina, is oversaturated. This causes, "merging" of areas of blood vessels and loss of detailed information. The blue channel (Figure 3d) is dominated by a uniform distribution of values over almost the entire area of the image, except for the area of the optic nerve disc and the overexposed area visible at the bottom of the image. This is due to the fact that the retina absorbs this component of the white light spectrum.

As a result of the numerical experiments, it turned out that the informations contained in the green and blue channels (which is the background in the source image) were the most useful. Based on the experiments, it was decided to eliminate the oversaturated red channel from the analyses. In addition, it turned out that the best results are obtained when the blue channel is attenuated (weighting of 0.4), while the green channel is taken in the full tonal dynamic range. An example of the intensity distribution of an image processed in this way, can be seen in the figure below (Figure 4).

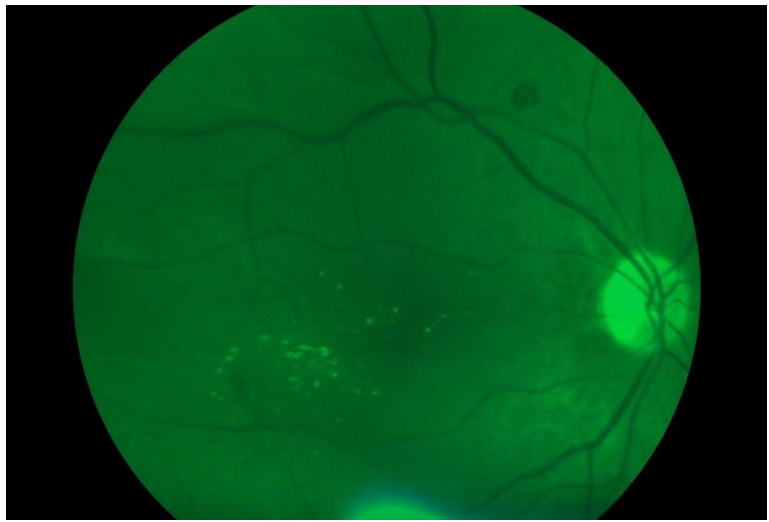

*Figure 4: Processed retinal image containing attenuated blue channel (weight 0.4) and green channel in full tonal dynamic range (weight 1.0), eliminated red channel.*

As part of the research work, a preliminary normalization of the images was also carried out, which included several consecutive steps. When defects such as rainbows, intense reflections, severely underexposed image, eyelashes in the image, etc. were identified, the retinal image was cropped from the top or bottom depending on the location of the defect.

Additional histogram equalization and pre-correction of saturation and contrast were intended to facilitate the selection of algorithm parameters to search for illness changes. On the other hand, blurring and subtraction of the averaged background of the image was intended to increase the contrast of the currently searched features. The analyses showed that for small values of the blur parameter, hard exudates and hemorrhages are best seen, while for large values of the blur parameter, the yellow macula could be easily found.

All the retinal images obtained were cropped to 90% of the length of their original radius. This step was performed to get rid of edge effects, including camera-generated descriptions and lighting inhomogeneities at the edges of the area captured by the camera. This step was done by defining the radius of the camera's registration area and then assigning zero values to the area outside this range (Figure 5).

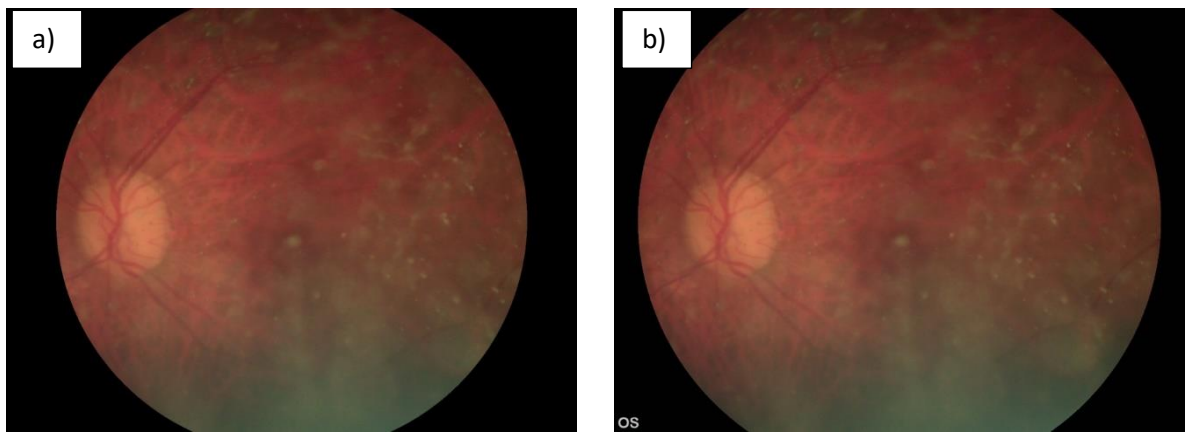

*Figure 5. Retinal image: a) original; b) cropped to 0.9 of the original diameter of the information area.*

Due to different lighting conditions, it is necessary to normalize the photos in terms of color saturation. In the case of images that have inhomogeneities resulting from overexposure of fragments with light coming from outside (insufficiently accurately applied camera cover) or eyelashes appearing in the field of view of the camera, it was necessary to remove these artifacts. Example retinal images subjected to the automatic normalization and cropping process are shown below (Figure 6).

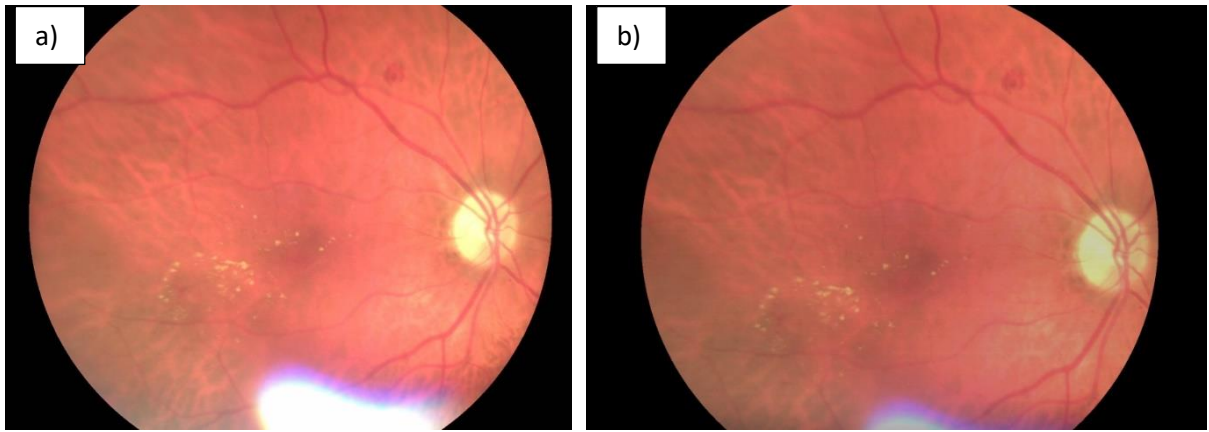

*Figure 6: Example of a retinal image - source image(a) and a processed, cropped and pre-normalized image (b).*

In some cases, automatic cropping of a photo led to a reduction of more than 30%. Such photos were rejected as unsuitable for further analysis. An example of such a photo is shown below (Figure 7).

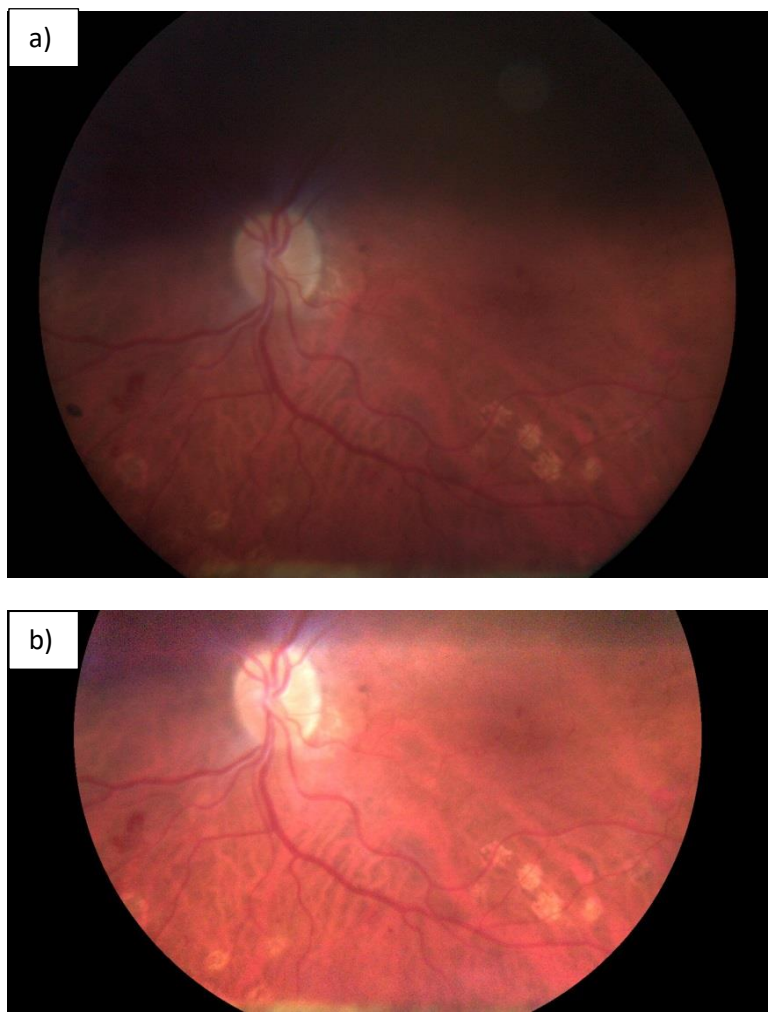

*Figure 7. Retinal photo: a) original; b) obtained after automatic cropping due to non-uniform illumination.*

Morphological algorithms defined in the OpenCV library were used to distinguish the optic nerve disc. Two algorithms in particular were used here, i.e. the dilation algorithm and the erosion algorithm. Morphological transformations are some simple operations based on the shape of the image. They are usually performed on binary images. They require two inputs, the first is the original image, and the second is the so-called structural element or kernel, which determines the nature of the operation. All morphological operations were carried out on images subjected to a pre-normalization process.

The figure below (Figure 8) shows the image presented in Figure 3a, which was processed using morphological algorithms. The figure (Figure 8a) shows the disappearance of detailed information in the image. In the figure, the position of the optic nerve disc (the circular area on the right side of the image) and the brightening in the lower part of the image due to the incorrect execution of the image (the camera hood not accurately attached to the orbit) are clearly visible, while the blood vessels in the fundus are not. The areas visible in the figure 8b were determined automatically and include the position of the optic nerve disc and the image defect visible in the lower part. Areas of the images, characterized by similar color and containing large clusters of hard exudates, can be similarly identified. In the figure, such exudates are visible (Figure 8b), but have not been marked due to their too low intensity.

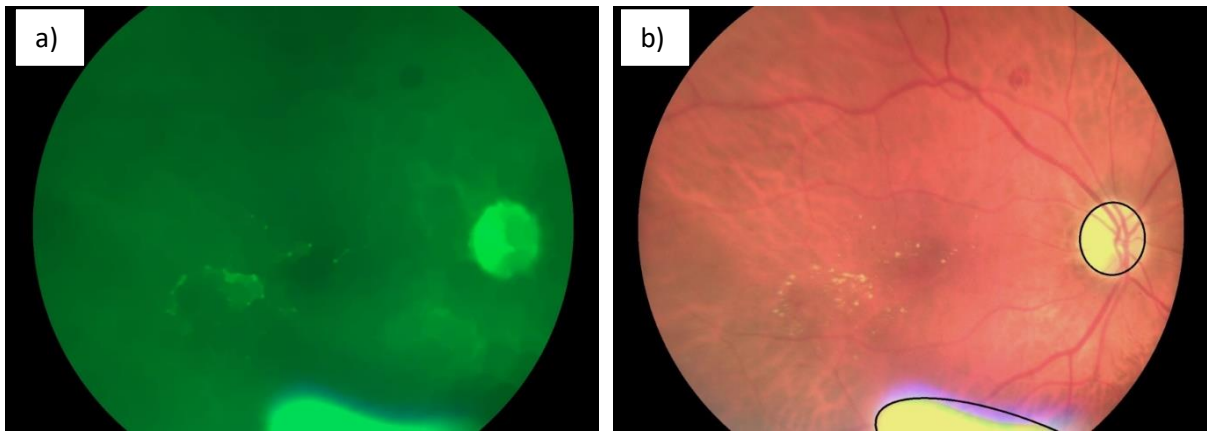

*Figure 8. a) Intensity distribution after morphological operations in the image shown in Figure 3a; b) original photo with the areas of optic disc and image defect marked.*

Due to the fact that the algorithm identified several areas characteristic of different structures, it is necessary to verify their origin. As a first step, it was decided to conduct an analysis to determine which area corresponds to the location of the optic nerve disc. The study took into account the fact that brightenings caused by improperly taken images are most often located near the borders of the image (this applies to images with correct geometry), while the optic nerve disc is usually the brightest element in the image. The morphological algorithms used were aimed at removing the blood vessel system, and the thresholding used was aimed at identifying the optic nerve disc. The algorithm's search of the image may have been disrupted by multiple clusters of hard exudates, the presence of which qualifies the patient for additional diagnostics. If they were identified, the resulting thresholding result was approximately coincident with the optic nerve disc found. The position of the optic nerve disc, determined by the algorithm, is also used to automatically reject images with abnormal geometry, such as when the optic nerve disc is in the center of the image, or is missing from the image. The yellow spot can be determined as the darkest part of the image from the previous step for large values of the blur parameter, or geometrically if the optic nerve disc is in the correct area of the photo. In the last step, the image was slightly brightened and it was checked that the determined position of the found disc did not change a lot, due to the presence of exudates that form large clusters. The next step of the algorithm's thresholding search for blood vessels. Pre-normalized images were subjected to thresholding to find dark areas. Since blood vessels have a similar hue to hemorrhages, they must be accurately classified. The classification process involved assessing the location of identified lesions within the blood vessel system, or checking whether the identified shape was approximately oval. Blood vessels were checked for their average width. For small threshold values, hemorrhages and blood vessel elements were mostly shown. Increasing the threshold value allowed the detection of a significant portion of the blood vessel system. The edges of the image were deliberately omitted from the analysis. In extreme cases, hard exudates have a much darker border, which can be mistaken during classification with hemorrhage. This does not change the fact that the proposed method of automatic diagnosis makes it possible to detect lesions and refer the patient to a specialized clinic.

The final thresholding stage of the algorithm was designed to detect hard lesions of the fundus. Characteristic areas of ocular elements identified in previous stages such as the optic nerve disc, yellow macula, blood vessels and classified hemorrhages were excluded in this thresholding stage. Thresholding was carried out starting with high threshold values in the search process, which made it possible to identify, as some of the first points on the diagram, lesions in the form of hard exudates. Characteristic of these lesions is their distribution and formation of clusters around the yellow spot, and thus their average distance was smaller. The rest of the image points are evenly distributed throughout the image. The most common overexposure and underexposure in the image made this step difficult, which in such cases required the use of local filters and algorithms for finding similarities, the effectiveness of which depends, among other things, on the contrast and overall quality of the fundus image. In the final stage of the work, the consistency of the algorithm's evaluations was tested using Cohen's kappa coefficient with quadratic weights. According to the literature, a kappa coefficient slightly greater than 0.8 is achieved by ophthalmology specialists [29]. The value of the kappa coefficient achieved by the algorithm for the data on which the algorithm was learned was 0.850.

A number of methods for processing and automatic analysis of retinal images can be found in the literature. In most works, their analysis is divided into stages that coincide with those used in the proposed approach. A distinction can be made between the stage of preprocessing and normalization of images, in which algorithms are used to remove inhomogeneous illumination and normalize and enhance contrast. The utility of applying these algorithms is considered in terms of improving the effectiveness of later algorithms performed on processed images. For example, a paper by Gnoheim [30] compared 7 normalization algorithms described in the literature for optimal determination of the distribution of blood vessels in the retina:
- Processing only the green channel (RGB),
- Histogram compensation,
- Adaptive local contrast enhancement method,
- Adaptive histogram equalization method,
- Desired average intensity method,
- Method of dividing by over-smoothed image,
- Subtraction by smoothed background image.

The authors indicated the method of adaptive histogram equalization [31]. The best accuracy of determining the distribution of blood vessels was adopted as the criterion.

The next step after preprocessing and normalizing the images is the stage of segmentation and localization of the optic disc and other retinal features. Methods for localizing and segmenting the optic disc (optic nerve disc) are divided by procedure into:

- Approaches based on intensity and shape features of the optic disc,
- Approaches using the location and orientation of the vascular system.

An example of the first approach is the algorithm presented in the work of Lalonde et al. [32]. In the first step, potential areas of the optic disc are located using pyramidal decomposition (an artificial reduction in image resolution) of the green channel in the RGB image. Through thresholding, the areas with the highest brightness at the lowest image resolution are found - usually the optic disc. This is followed by an operation to find edges in the image and several more mathematical operations to adjust the circular template. As a result of such an operation, the authors obtained precisely defined boundaries of the optic disc. Algorithms based on the second approach are based on the fact that there are many blood vessels leading to the optic disc. By studying the directions of their propagation and density, it is possible to determine the location of the optic disc [33]-[35]. Also, the yellow macula can be found based on geometric relationships. It is located at a distance of 2.5 optic nerve disc diameters from the center of the disc, between the main arms of the circulatory system. An algorithm based on this assumption is presented in the work of Li and Chutatape [36].

Also, exudate detection techniques can be divided into two groups. Algorithms in the first group are based on mathematical morphology, while those in the second are based on pixel classification. In the work of Walter et al. [37] the authors used morphological closure as the first step to eliminate the vessel. Next, the local standard deviation was calculated to isolate potential exudates. Finally, a morphological reconstruction method was used to find the contours of the structures. In retinal images, similar brightness to exudates is often found in the optic disc, so in the work of Sophoraki et al. [38] eliminated the optic disc as early as the first step. They then used Otsu binarization to localize high-intensity regions. In the work of Habib et al. [39] an initial set of potential exudates was determined using a matched Gauss filter. To reduce false indications, a classifier was used with a set of 70 features most commonly found in the literature. With appropriate modifications, the above methods can also be used to segment hemorrhages [38], [40].

The final step was to create an application with a graphical user interface (Figure 9) that allowed retinal images taken at cooperating ophthalmology offices to be uploaded to the server. These images were then analyzed using a developed algorithm to make a diagnosis.

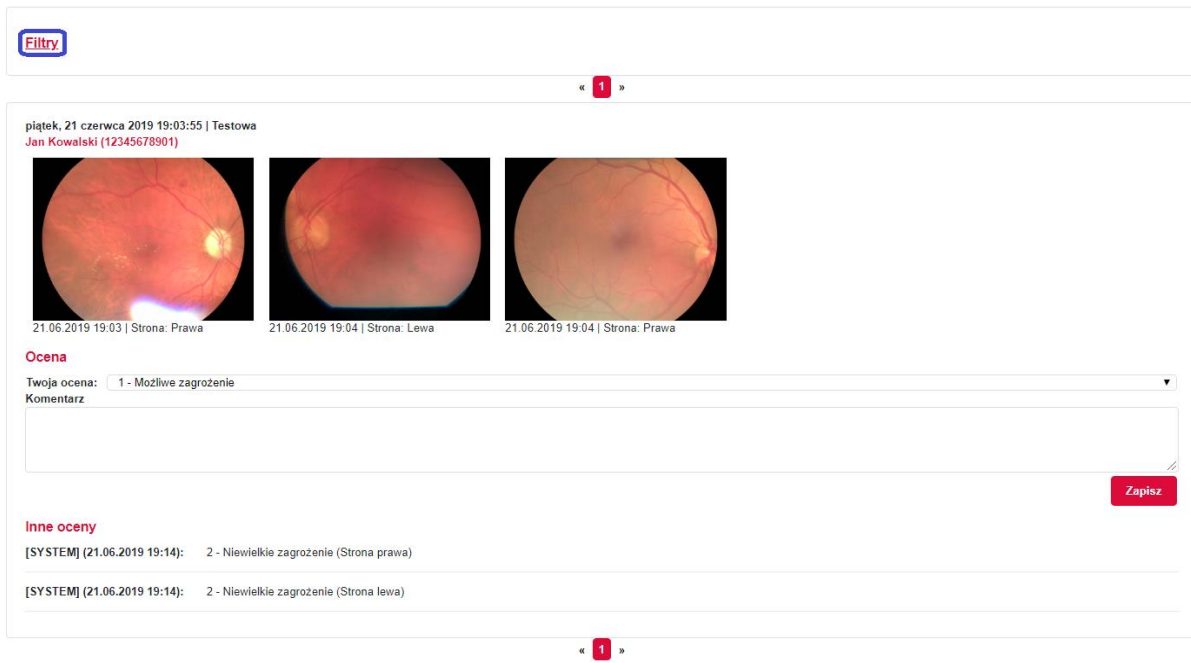

*Figure 9. Diabetic retinopathy diagnostic application window with patient test images loaded*

**Conclusions**

The present research focuses on the automatic analysis and normalization of patients retinal images obtained using a portable ophthalmoscope, which are a key step in implementing an algorithm for diagnosing diabetic retinopathy.

The examined retinal images were featured by varying quality and numerous defects, which make accurate analysis of the images much more difficult, which was the main challenge in the application of automatic diagnostic techniques. A number of steps were carried out to improve the quality of the images, including the elimination of low-quality images and initial normalization. These steps helped remove defects and increase contrast in the images, which significantly affected the quality of the analysis. The conducted color channel analysis proved that the most important information is contained in the green and blue channels, while the red channel could be eliminated from the analysis. This approach contributed to increasing the efficiency of the algorithm.

Subsequent steps such as segmentation and localization of features, such as the optic nerve disc, were crucial in analyzing the images. The morphological and thresholding algorithms used enabled precise localization of these structures.

Verification of the algorithm showed its high agreement with the assessments of ophthalmology specialists, which proves its usefulness in the automatic diagnosis of diabetic retinopathy.

The conclusions of the study and the developed algorithm are important in speeding up the process of diagnosing diabetic retinopathy and may find application in other fields of medicine. This work contributes valuable information on retinal image processing, which may be beneficial for the development of similar diagnostic systems.

**Acknowledgements**

The research presented in this paper was financed by the European Regional Development Fund under the project „Badania nad nowymi metodami diagnostycznymi wczesnego wykrywania retinopatii cukrzycowej", implemented under the program Regional Operational Program of the Podkarpackie Voivodeship 2014-2020, Priority Axis: I Competitive and innovative economy, Action 1.2 Industrial research, development works and their implementation, grant agreement number: RPPK 01.02.00-00-18-015/17-00.